\documentclass[onecollarge,natbib]{svjour2}
\bibpunct{[}{]}{;}{n}{}{,} % to get "[numbered]" references from natbib
\smartqed  % flush right qed marks, e.g. at end of proof
\usepackage{graphicx}
\usepackage{amsbsy}
\usepackage{amsmath}
\usepackage{amsfonts}
\journalname{Few-Body Systems}
\begin{document}

\title{Efimov physics with $1/2$ spin-isospin fermions}

\author{A. Kievsky and M. Gattobigio}

\institute{A. Kievsky \at
Istituto Nazionale di Fisica Nucleare, Largo Pontecorvo 3, 56100 Pisa, Italy \\
  \email{alejandro.kievsky@pi.infn.it}           %  \\
           \and
M. Gattobigio \at
  Universit\'e de Nice-Sophia Antipolis, Institut Non-Lin\'eaire de
    Nice,  CNRS\\ 1361 route des Lucioles, 06560 Valbonne, France \\
  \email{mario.gattobigio@inln.cnrs.fr}           %  \\
}

\date{Received: date / Accepted: date}

\maketitle

\begin{abstract}
The structure of few-fermion systems having $1/2$ spin-isospin symmetry
is studied using potential models. The strength and range of the two-body potentials
are fixed to describe low energy observables in the angular momentum $L=0$ state
and spin $S=0,1$ channels of the two-body system.
Successively the strength of the potentials are varied in order to explore energy regions
in which the two-body scattering lengths are close to the unitary limit. This study
is motivated by the fact that in the nuclear system the singlet and triplet scattering
lengths are both large with respect to the range of the interaction. Accordingly we
expect evidence of universal behavior in the three- and four-nucleon systems that can be 
observed from the study of correlations between observables. In particular we concentrate
in the behavior of the first excited state of the three-nucleon system as the system
moves away from the unitary limit. We also analyze the dependence on the range of the three-body
force of some low-energy observables in the three- and four-nucleon systems. 
\end{abstract}

\section{Introduction}\label{sec:introduction}

The study of few-body systems in the limit of large two-body scattering length $a$
is an intense subject of research nowadays. In this limit the systems show
universal behavior largely independent of the range and characteristic of the particular
interaction. Research on universal behavior in few-body systems has been triggered by 
V. Efimov in a series of papers~\cite{efimov1,efimov2}. 
He demonstrated that in the unitary limit, $1/a=0$, the three-boson system has an infinite
number of excited states that accumulate to zero following a geometrical series.
As the scattering length moves from the unitary limit to finite values, the
infinite energy levels become finite tending to the natural distribution of
levels observed in most real systems. The very particular energy spectrum of a three-boson
system at the unitary limit is called Efimov effect and the study of the dynamics
of such a system close to the unitary limit is called Efimov physics (for a recent review
see Ref.~\cite{report}).

The theory developed by Efimov is a zero-range theory (scaling limit). In this limit
the two-body energy is $E_2=\hbar^2/ma^2$ whereas in real systems 
the energy scattering length $a_B$ defined from the two-body energy 
as $E_2=\hbar^2/ma_B^2$ could be very different from $a$. Efimov physics describes systems
in which $a$ and $a_B$ are of the same order, both being very large with respect to the range
of the interaction $r_0$. To this respect a lot of efforts have been given to incorporate
range corrections in the zero-range theory
(see for example Refs.~\cite{platter2009,gatto2014}). 
In fact, experiments devoted to study Efimov 
physics~\cite{ferlaino2011,machtey2012,roy2013,dyke2013} show some discrepancies when the results are
compared to the zero-range theory. In order to clarify in which form the range corrections
can be taken into account we recall the Efimov radial law which describes the three-boson
energy spectrum as a function of the two-body scattering length $a$. In parametric form
it can be expressed as,

\begin{equation}
  \begin{gathered}
    E_3^n/(\hbar^2/m a^2) = \tan^2\xi \\
    \kappa_*a = \text{e}^{(n-n^*)\pi/s_0} 
    \frac{\text{e}^{-\Delta(\xi)/2s_0}}{\cos\xi}\,.
  \end{gathered}
  \label{eq:energyzr}
\end{equation}
where $\kappa_*$ is the three-body parameter that fixes the energy of the
three-boson system, $\hbar^2 \kappa_*^2/m$, for the level $n=n^*$, at the unitary limit.
In the above equation $a$ is the control parameter and, once the value of the
three-body parameter is fixed, a complete knowledge of the spectrum requires
the knowledge of the universal function $\Delta(\xi)$. The above equation
shows a discrete scale invariance (DSI), the energy ratio of two consecutive states is
$E_3^n/E_3^{n+1}=e^{2\pi/s_0}\approx 515.03$, with $s_0\approx 1.00624$ a universal
number, at fixed values of the angle $\xi$. This property is a consequence of the fact
that the universal function $\Delta(\xi)$ is the same for all levels. However it should be
pointed out that the spectrum given by Eq.(\ref{eq:energyzr}) is unbounded from below,
this is known as the Thomas collapse~\cite{thomas:1935_phys.rev.}. In order to compute
$\Delta(\xi)$ some regularization has to be introduced. In practice the 
Skorniakov-Ter-Martirosian (STM) equation has been solved with some cutoff and $\Delta(\xi)$
has been obtained analyzing the energy spectrum of the first excited states minimizing
the cutoff effects~\cite{hammer2003}. A parametrization of the universal function can be
found in Ref.~\cite{report}.

Instead of solving the STM with a cutoff, the three-boson system close to the unitary limit
can be studied using potential models. This has been done by the authors in a series of
papers~\cite{gatto2014,kievsky2013,gatto2012,kiev2014}
in which a Gaussian potential with variable strength has been used. The conclusion was
that the Efimov radial law can be modified as
\begin{subequations}
  \begin{eqnarray}
  \label{eq:energyfrA}
    E_3^n/E_2 = \tan^2\xi \\
    \kappa_na_B + \Gamma_n =
    \frac{\text{e}^{-\Delta(\xi)/2s_0}}{\cos\xi} \,.
  \label{eq:energyfrB}
  \end{eqnarray}
  \label{eq:energyfr}
\end{subequations}
where the energy scattering length $a_B$ is defined as $E_2=\hbar^2/ma_B^2$. 
For positive values of $a$ $E_2$ is the two-body binding energy
whereas for negative values of $a$
$E_2$ is the energy of the two-body virtual state. Moreover $\kappa_n$ defines
the energy of level $n$ at the unitary limit, $\hbar^2 \kappa_n^2/m$. 
Besides the modification at the two-body level, a finite-range parameter, $\Gamma_n$, has been
introduced as a shift in the control quantity $\kappa_na_B$. This form of the radial law has
been justified in the context of effective field theory introducing an Efimov running
parameter~\cite{ji:2015}. In a recent discussion the authors have shown that the ground 
and first excited state shifts, $\Gamma_0$ and $\Gamma_1$, have an almost constant behavior 
close to the unitary limit~\cite{kievsky2015}.

The success of using Eq.(\ref{eq:energyfr}) to describe the ground and first excited states of three 
bosons suggests the possibility of extending the analysis to systems having spin-isospin symmetry
as the three-nucleon system. As it is well known, this system has a large probability of
being in the symmetric $L=0$ state. Moreover the singlet and triplet $np$ scattering lengths,
$a^0$ and $a^1$, are both large with respect to the range of the nuclear interaction.
Accordingly in the present work we make a detailed analysis of the Efimov
physics for three $1/2$ spin-isospin fermions around the unitary limit. Preliminary studies
on this subject can be find for example in Ref.~\cite{efimov3}. Our analysis is based on the
solution of the Schr\"odinger equation using a potential model 
constructed to describe some low energy observables in the two-nucleon system. 
To this aim we use a spin dependent Gaussian potential
with the ranges and strengths fixed to describe the scattering lengths and some other
quantities as the effective ranges in singlet and triplet state and the deuteron binding
energy. After this construction, the exploration of the physics
close to the unitary limit is done by the variation of the Gaussian strengths in both channels
at constant range values. This will be done for the two- and three-fermion systems.
When the values
of the scattering lengths are equal to the experimental values, the analysis can be
used to compare the quality of this description for the low-energy observables to the experimental
data as the triton binding energy and the low energy $nd$ phases 
or, extending the analysis to the four-body system, the predictions on the $\alpha$ particle 
binding energy. As we will see the inclusion of a three-body force
allows for a quantitative description of the low energy observables.

The paper is organized as follow. In section 2 the two-body system is discussed.
The Gaussian potentials in the two spin channels are introduced and the parameters used to fit
the strengths and ranges are indicated. The behavior close to the 
unitary limit is analyzed. In section 3 the three-fermion system is studied 
along planes at fixed values of the ratio of the singlet and triplet scattering lengths. 
The energy spectrum has been computed and the behavior of the first excited state
has been followed as the scattering length values move away from the unitary limit.
In section 4 we introduce the three-body force an analyze its dependence on
specific observables as the doublet $n-d$ scattering length and the $^4$He ground state. 
In section 5 we discuss the results and perspectives of the present study. 

\section{Two $1/2$ spin-isospin fermions close to the unitary limit}\label{sec:twof}

In the following we consider a system of identical fermions having $1/2$ spin-isospin
symmetry. Having in mind the nuclear system, we consider that the fermions can interact
differently in the spatially symmetric $L=0$ state with spin-isospin 
$S=0,T=1$ or $S=1,T=0$. These two states can be therefore characterized by two different 
scattering lengths in the singlet state, $a^0$, and in the triplet state,
$a^1$. The case we are interested here is the case in which both
lengths are large with respect to the range of the two-fermion interaction. 
In the case in which one of the scattering lengths is positive the two-body
system supports a shallow bound state whereas for negative values the two-body
system has a shallow virtual state. The energy of the bound or virtual state characterizes
the state too. Therefore there are two quantities, the energy of the shallow state
and the scattering length characterizing the two-fermion system in each of the two
spin states. We can use this minimal information to construct a spin dependent
two-parameter potential to describe the two-fermion system in the $L=0$ state. 
Then this description can be extended to describe the three- and four-fermion 
systems. 

Following previous works in identical boson systems we consider a Gaussian potential
of the form
\begin{equation}
V(r)=V_S {\rm e}^{-r^2/r_S^2}\, ,
\label{eq:gauspot}
\end{equation}
in which the strength $V_S$ and the range $r_S$ can depend on the spin $S$ of the
state. In the case of the $n-p$ system, the experimental data 
characterizing the spin-isospin $S=1,T=0$ state are the deuteron energy of $-2.225$ MeV and
the triplet scattering length $a^1=5.419\pm0.007\;$fm. In the case of the singlet state the
scattering length results $a^0=-23.740\pm0.020\;$fm. These quantities can be supplemented
by the effective range in each spin state, $r_{eff}^S$ determined from the low-energy
$n-p$ scattering data. Their values are $r_{eff}^0=2.77\pm0.05\;$fm and
$r_{eff}^1=1.753\pm0.008\;$ fm respectively for $S=0,1$
(for the latest experimental values of these quantities see Ref.~\cite{report2011} and
references therein).

In order to see the capability of the Gaussian potential defined in Eq.(\ref{eq:gauspot}) 
to reproduce the experimental quantities, in Table~\ref{tab:tab1} we report the
corresponding estimates for selected values of $V_S$ and $r_S$.

\begin{table}[h]
\begin{tabular}{llll|lllll}
 $V_0$[MeV] & $r_0$[fm] & $a^0$[fm] & $r^0_{eff}$[fm] &
 $V_1$[MeV] & $r_1$[fm] & $a^1$[fm] & $r^1_{eff}$[fm] & $E_2$[MeV] \\
\hline
 $-53.255$ & $1.40$ & $-23.741$ & $2.094$ & $-79.600$ & $1.40$ & $5.309$ & $1.622$ &-2.225 \\
 $-42.028$ & $1.57$ & $-23.745$ & $2.360$ & $-65.740$ & $1.57$ & $5.425$ & $1.776$ &-2.225 \\
 $-40.413$ & $1.60$ & $-23.745$ & $2.407$ & $-63.712$ & $1.60$ & $5.447$ & $1.802$ &-2.225 \\
 $-37.900$ & $1.65$ & $-23.601$ & $2.487$ & $-60.575$ & $1.65$ & $5.482$ & $1.846$ &-2.225 \\
 $-33.559$ & $1.75$ & $-23.745$ & $2.644$ & $-55.036$ & $1.75$ & $5.548$ & $1.930$ &-2.225 \\
 $-30.932$ & $1.82$ & $-23.746$ & $2.756$ &          &        &         &         & \\
\hline
\end{tabular}
\caption{Scattering length $a^S$ and effective range $r^S_{eff}$ for selected values of the
strength $V_S$ and range $r_S$ of the Gaussian potential in states with spin $S=0,1$.
For the triplet state the two-body binding energy, $E_2$, is also shown.}
\label{tab:tab1}
\end{table}

From the table we can observe a good description of the experimental values with the best
description obtained using different ranges in the singlet and triplet states. For example,
the parameters on the last row for $S=0$ and in the second row for $S=1$ give a very
good description of the experimental results. The parameters on the fourth row give
the best description considering the same range in both spin states.
In the following we make an analysis of the two-fermion system close to the unitary
limit using the same range, $r_0=r_1=1.65\;$fm, for the singlet and triplet
states. 

\begin{figure}[h]
\vspace{1.2cm}
\begin{center}
\includegraphics[width=\linewidth]{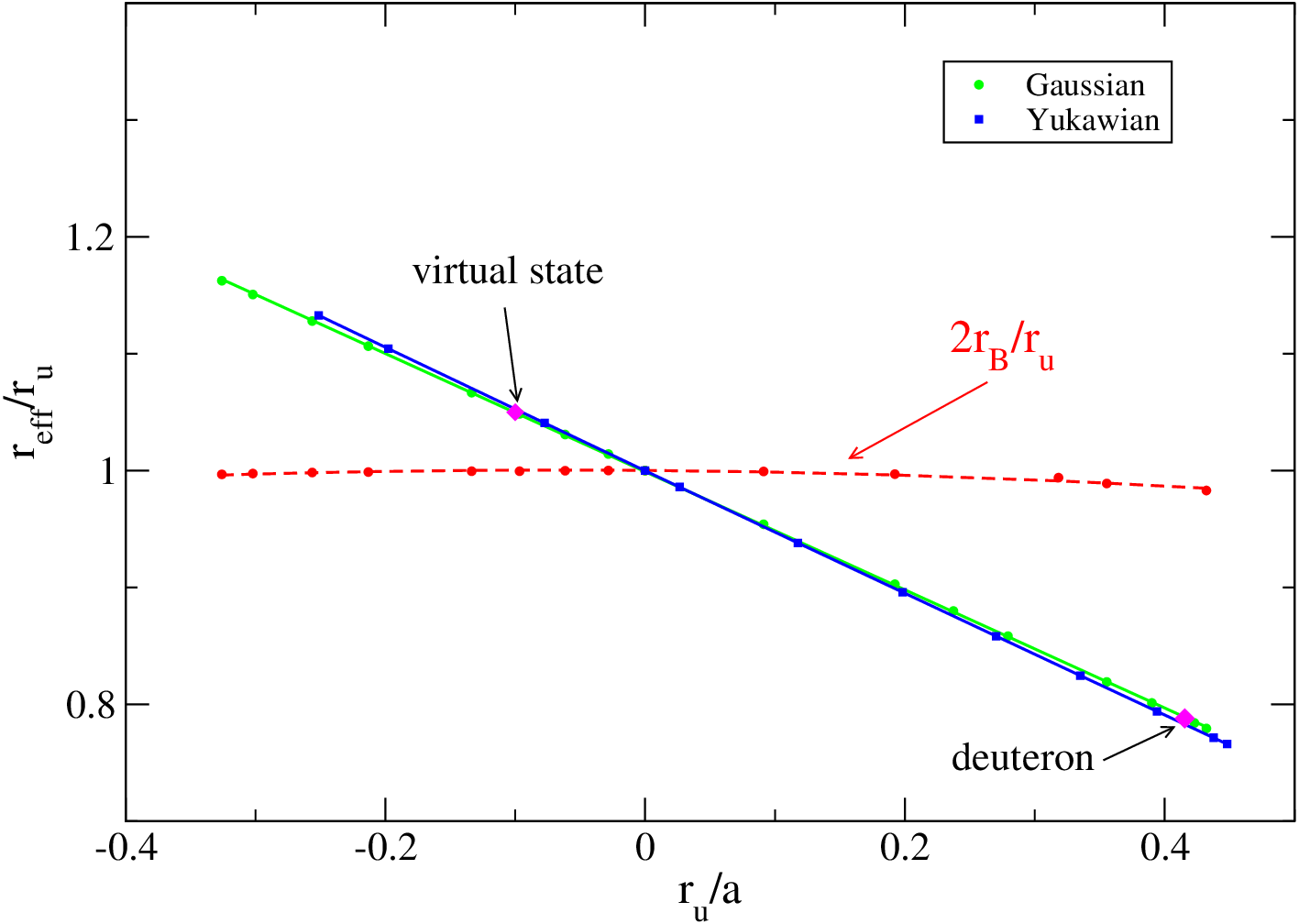}
\end{center}
\caption{(color online). The effective range $r_\text{eff}$ and the length $2r_B$ 
as a function of the inverse of the scattering length (in units of $r_u$)
for the Gaussian potential. The deuteron and virtual state points are shown. For
the sake of comparison results obtained with a yukawian potential are given too.}
\label{fig:rb}
\end{figure}

In order to explore the dynamics of the system close to the unitary
limit we allow variations in the strength of the Gaussian potential covering 
values of $a^S$ that verify $a^S>r_0$,
eventually the unitary limit, $1/a^S=0$, can be reached. 
The variation of the potential strength defines a particular
path to connect the physical points to the unitary limit. Along this path the
scattering length $a_S$ and the energy scattering length $a_B^S$ are related
by the effective range expansion
\begin{equation}
   \frac{1}{a_B^S}\approx \frac{1}{a^S} + \frac{r^S_{eff}}{2(a_B^S)^2} \, .
\label{eq:reffexp}
\end{equation}
The above relation is approximate since higher order terms are present
as the two-body energy increases. However it is very well fulfilled
close to the unitary limit. Moreover it reflects the fact that 
in the zero-range limit (scaling limit) the scattering length
and the energy scattering length are equal. Eq.(\ref{eq:reffexp}) can be
put in the form
\begin{equation}
   r^S_{eff}a^S=2 a_B^S r_B^S
\label{eq:reffaB}
\end{equation}
where we have introduced the length $r_B^S=a^S-a_B^S$. The quality of the above
relation can be checked for example at the physical points using the values 
given in Table~\ref{tab:tab1}. We consider the $r_0=r_1=1.65\;$ fm case and therefore,
for $S=0$ taking the Gaussian strength $V_0=-37.90\;$MeV, 
the energy of the virtual state results
$E_2=-0.0675\;$MeV and the above relation is verified at the level of $0.2\%$.
For the $S=1$  the relation is verified at the level
of $0.6\%$. Moreover, it can be observed that in this two cases the length $r_B^S$
remains almost constant; the values are $r^0_B=1.183\;$fm in the first case and 
$r^1_B=1.165\;$fm in
the second case. To analyze further this fact we can cast Eq.(\ref{eq:reffexp})
in the generic form

\begin{equation}
  \frac{r_{eff}}{2r_B}=1-\frac{r_B}{a} + {\cal O}(1/a^2)\, ,
  \label{eq:rbvsre}
\end{equation}
where we have dropped the superscript $S$ and we have explicitly stated that
there are quadratic corrections to the formula. If the length $r_B$ were 
constant we could replace it by its value at the unitary limit, $r_u/2$ with
$r_u$ the effective range at this limit. In this case Eq.(\ref{eq:rbvsre})
expresses the following universal relation 
\begin{equation}
  \frac{r_{eff}}{r_u}=1-0.5 \frac{r_u}{a} 
  \label{eq:reffun}
\end{equation}
that can be studied using the Gaussian potential. This is done in Fig.~\ref{fig:rb} 
in which the quantities $r_{eff}/r_u$ and $2r_B/r_u$ are plotted as a function
of $r_u/a$ for positive and negative values. The points corresponding to the
virtual state and deuteron state are explicitly shown.
From the figure we can see that the results using the Gaussian potential with variable
strength (given by the green solid circles) connects linearly these two points going 
through the unitary limit. The solid (green) line is a fit of the results using
Eq.(\ref{eq:reffun}).
Along this path the length $r_B$ is almost constant as is given by the
(red) solid points. The (red) dashed line is a fit to the numerical
results using a linear plus a very small quadratic $r_u/a$ term.
In order to see the quality of the description of the Gaussian
potential, we have analyzed a more realistic potential constructed to
describe the two-nucleon system in the low energy region, 
the MTIII potential~\cite{mtjon}. This interaction consists in
a sum of two yukawians and it was used many times in the literature
to describe few-nucleon systems. Here we have varied the
strength of the interaction in order to cover the same range of
scattering length values analyzed with the Gaussian potential. The
results in this case, shown in the figure as (blue) squares, are very
close to the Gaussian results. The solid (blue) line is a fit to the results using
the form given in Eq.(\ref{eq:reffun}). We can conclude that close to the
unitary limit a two-parameter potential, as the Gaussian interaction,
contains the essential ingredients to describe the two-body dynamics.

\section{Three $1/2$ spin-isospin fermions close to the unitary limit}\label{sec:threef}

The Gaussian potential with variable strength in the two spin channels,
$S=0,1$, can be used to study the three-fermion spectrum. As in the case of 
bosons we can study the energy of the system as a function of the
(inverse) of the scattering lengths in order to construct the Efimov plot. 
It should be noticed that in the present case there are two scattering lengths
and so, the usual two-dimensional plot, which represents the three-body energy
as a function of the inverse of the scattering lengths, becomes three-dimensional.
Among different possibilities it is possible to define planes for fixed values
of the ratio $a^0/a^1$ and to study the energy
of the system along the planes defined by those ratios. This means that 
varying the $S=1$ Gaussian strength in order to have a large $a^1$ value, the
$S=0$ Gaussian strength will be varied accordingly to reproduce the selected ratio.

\begin{figure}[h]
\vspace{1.2cm}
\begin{center}
\includegraphics[width=\linewidth]{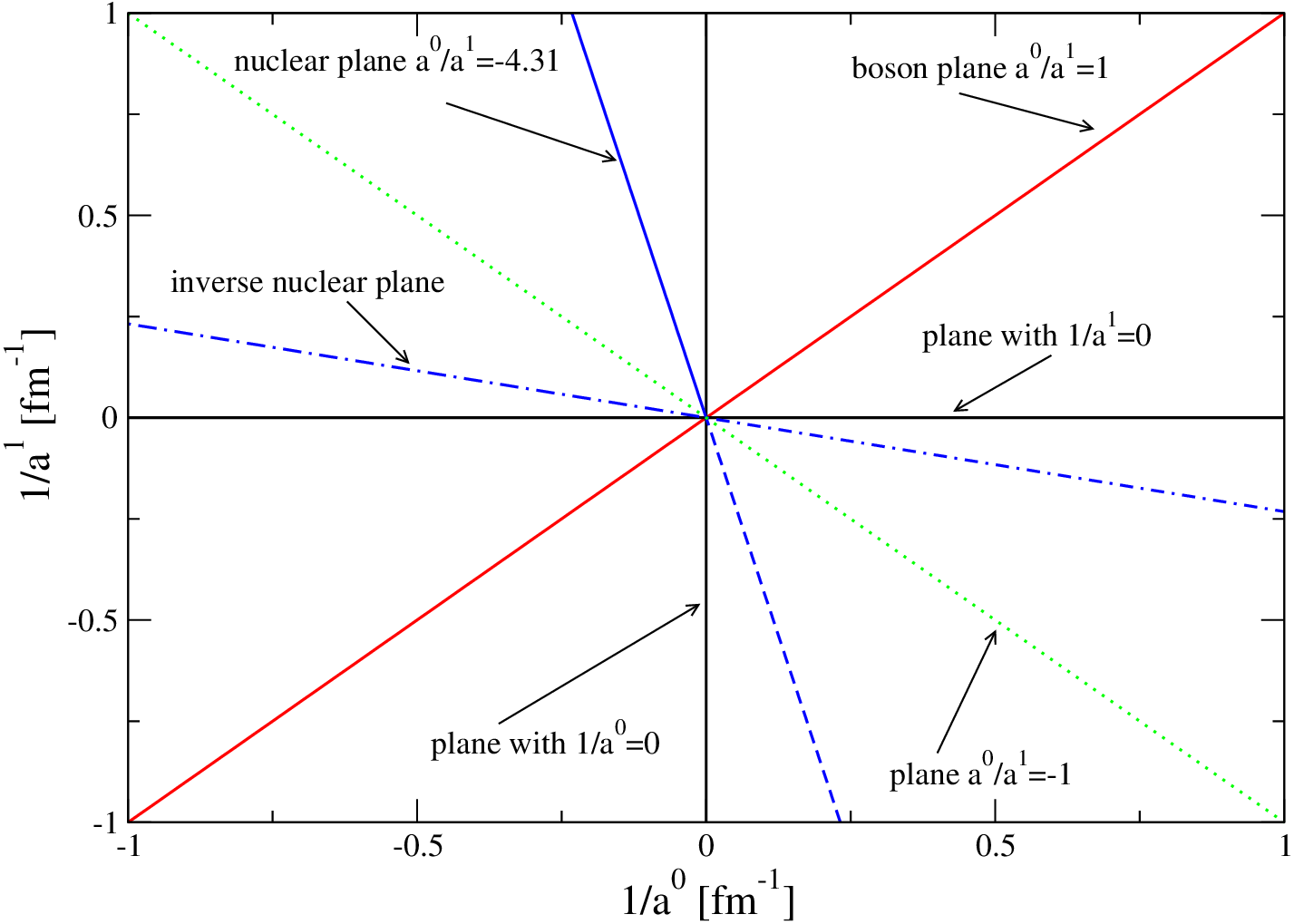}
\end{center}
\caption{(color online). Different planes are shown for specific values of the
singlet and triplet scattering lengths $a_0$ and $a_1$.}
\label{fig:planes}
\end{figure}

Example of planes at fixed values of the $a^0/a^1$ ratio are shown in Fig.~\ref{fig:planes}.
As mentioned in the previous section, the present study is done with Gaussian of equal
range. In this case, when the two scattering lengths are equal, the plane corresponds
to the bosonic case as it is indicated in the figure with the (red) solid  line. 
Another important
case is when the Gaussians have the strengths defined in the fourth row of 
Table~\ref{tab:tab1}, in this case $a^0/a^1=-4.31$ defining a plane in which the 
nuclear physics point is included. We define this plane the nuclear physics plane
and it is shown in the figure with a (blue) solid line located in the second quadrant.
Other special cases are shown in the figure as well. In the figure some symmetries
can be identified. When the potentials in the two spin channels are exchanged the
energy of the system does not change and the spectrum is the same. Accordingly there
is a reflection symmetry along the boson plane. In the case of the second and
fourth quadrant there is a reflection symmetry along the $a^0/a^1=-1$ plane.
These symmetries reduced the sectors where studying the three-fermion spectrum.

\begin{figure}[h]
\vspace{1.2cm}
\begin{center}
\includegraphics[width=\linewidth]{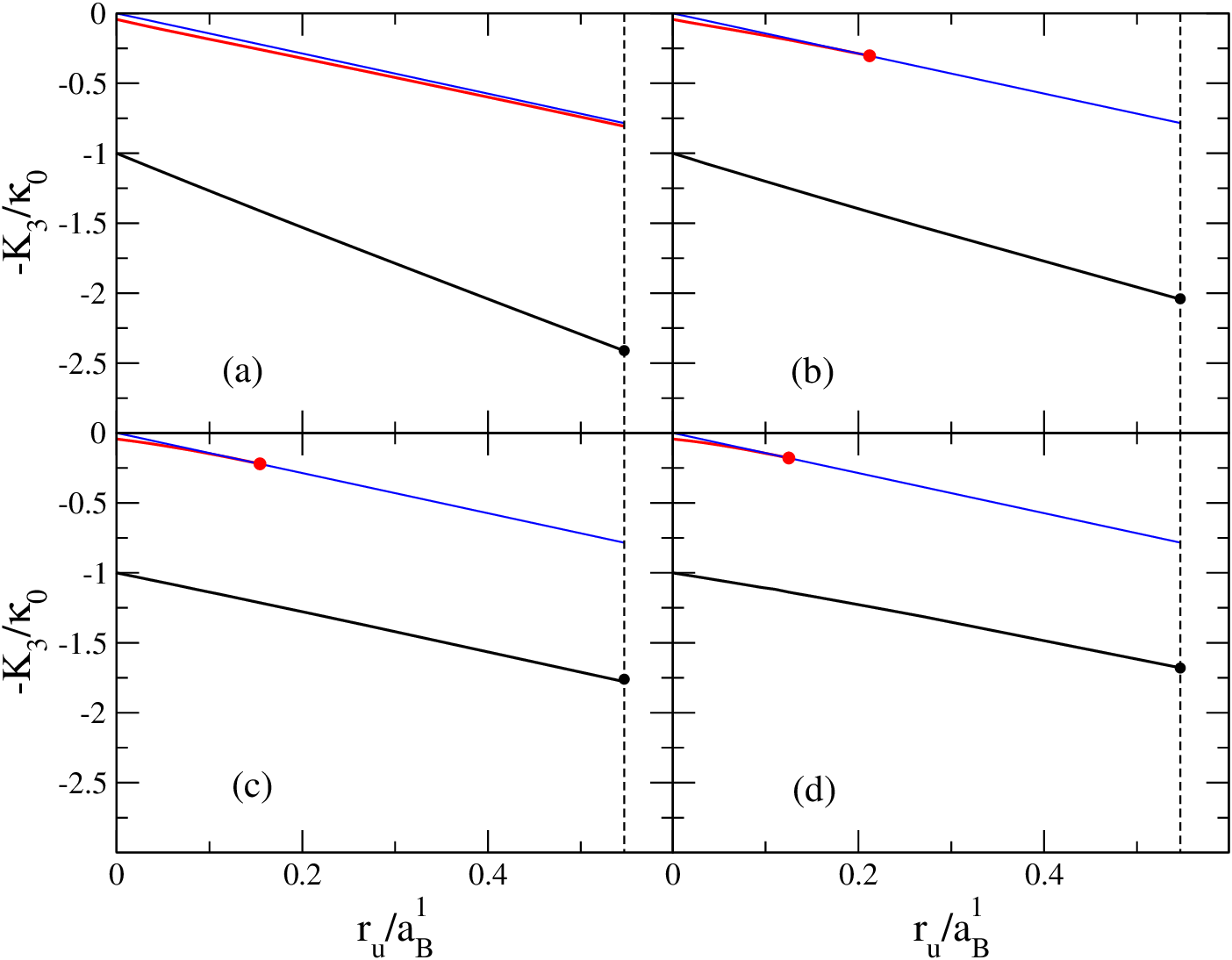}
\end{center}
\caption{(color online). The binding momentum $K_3$ (in units of $\kappa_0$) for the ground state
(black) thick line and first excited state (red) thick line as a function of
the inverse of triplet scattering length $a^1$ (in units of $r_u$). The (blue) thin line is the
two-body binding momentum. The singlet scattering
length $a^0$ is fixed by the ratio $a^0/a^1=1$ (a), $a^0/a^1=2$ (b), $a^0/a^1=\infty$ (c)
and $a^0/a^1=-4.31$ (d). The solid (red) circle indicates the position of the excited
state at the two-body threshold. The vertical dashed line indicated the physical
point and the solid (black) circle the value of the ground state binding momentum at that
point.}
\label{fig:efimov}
\end{figure}

Using the Gaussian spin dependent potentials with strength parameters fixed
to give scattering length values
along the planes discussed above, we study the $L=0$ three-fermion spectrum in the
$J^\pi=1/2^+$ state. To this aim we solve the Schr\"odinger
equation for different values of the Gaussian potential strengths in order
to cover a wide range of scattering length values. In this particular study we
have fixed $r_0=r_1=1.65\;$ fm. The results, following
planes as those defined in Fig.~\ref{fig:planes}, are collected in Fig.~\ref{fig:efimov}
for four selected cases. Positive values of the triplet energy scattering length $a_B^1$ have
been considered. In the figure the
binding momentum, $K_3$, defined from the three-body energy $E_3=\hbar^2 K_3^2/m$,
is shown for the ground and first excited state as a function of $a_B^1$ 
(in units of $r_u$). The ground state energy at the unitary limit, $\hbar^2\kappa_0^2/m=3.62$ MeV,
is the same in all cases; it corresponds to tune both scattering lengths at
the unitary limit, $1/a^0=1/a^1=0$. The binding momentum is given in units of $\kappa_0$. 
In each panel the ratio $a^0/a^1$ has been
kept constant. Panel (a) and (b) correspond to $a^0/a^1=1$ and $2$ respectively
whereas in panel (c) the Gaussian strength in the singlet state is kept
fixed at the unitary limit, therefore $1/a^0=0$ and $a^0/a^1=\infty$. 
Finally in panel (d) we consider
the ratio $a^0/a^1=-4.31$ containing the physical point. In the four panels the
thin (blue) line corresponds to the two-body binding energy whereas the thick
(black) and (red) lines correspond to the ground state and first excited state
energy, $E_3^0$ and $E_3^1$, respectively. 

At the unitary limit the spectrum shows
the Efimov effect, a geometrical series of excited states appear. Due to the very different scales
of such states, in Fig.~\ref{fig:efimov} only the ground and first excited states are shown.
In the zero-range
limit the geometrical factor defining the ratio between two consecutive energy state
is $\approx 515.03$ for all the states due to the DSI. Using the Gaussian potential
the ratio between the ground and first excited state at the unitary limit
is $\approx 528$ showing some range corrections. For the Gaussian range considered of $1.65\;$fm
the excited state energy at the unitary limit is $E_3^1=0.00685$ MeV.
As $a_B^1$ moves away from the unitary limit towards lower (positive) values, the higher excited 
states disappear into the nucleon-deuteron threshold.
It is possible that for specific values
of $a_B^1$ also the first excited state $E_3^1$ disappear into the two-body continuum.
This happens for the cases shown in panels (b), (c) and (d) at the point indicated by 
the (red) solid circle whereas in panel (a), describing a three-boson
system, $E_3^1$ is always below the two-body threshold given by $E_2$. 
The vertical dashed lines in the panels correspond to the deuteron binding energy and the (black)
solid circle indicates the ground state energy of the three-fermion system at that point. 
The values are
$E_3^0=-21.05$ MeV,$-15.12$ MeV, $-11.44$ MeV and $-10.22$ MeV for panels (a), (b), (c) and
(d) respectively. 

In panel (d) the ratio of the scattering lengths corresponds approximately to the experimental
ratio. Therefore the three-body energy $E_3^0=-10.22$ MeV, calculated at the value of 
$a^1_B\approx 4.32\;$fm can be considered an estimate of the $^3$H energy. 
However, from the study in boson systems, it is
well known that in order to correctly describe the three-body energy with the minimal
information introduced in the two-body potentials, a three-body force has to
be included~\cite{gatto2014,kievsky2013,bedaque99}. In the present case we observe an overbinding
of the $^3$H binding energy compared to the experimental value of $-8.48$ MeV, which is
usual in this type of description. Accordingly a repulsive three-body force is needed 
for a quantitative description of the three-nucleon system.
The discussion of the system including a three-nucleon force is given in the next section. 
Another important conclusion of the results
shown in Fig.~\ref{fig:efimov} is the behavior of the first excited state energy $E^1_3$. For the
boson system, given in panel (a), a two level system is observed at all values of 
$a_B^1$. In the other planes with bigger values of the ratio $a^0/a^1$
this state disappear into the two-body continuum at certain values of $a^1$. 
In particular, in panel (d) it is not present at the physical value
of $a_B^1$, in agreement with the experimental situation. This study agrees with previous analyses
on the first excited state and its behavior as a virtual state as it moves into the two-body
continuum~\cite{adhikari1982,frederico1988}.

\begin{figure}[h]
\vspace{2.5cm}
\begin{center}
\includegraphics[width=\linewidth]{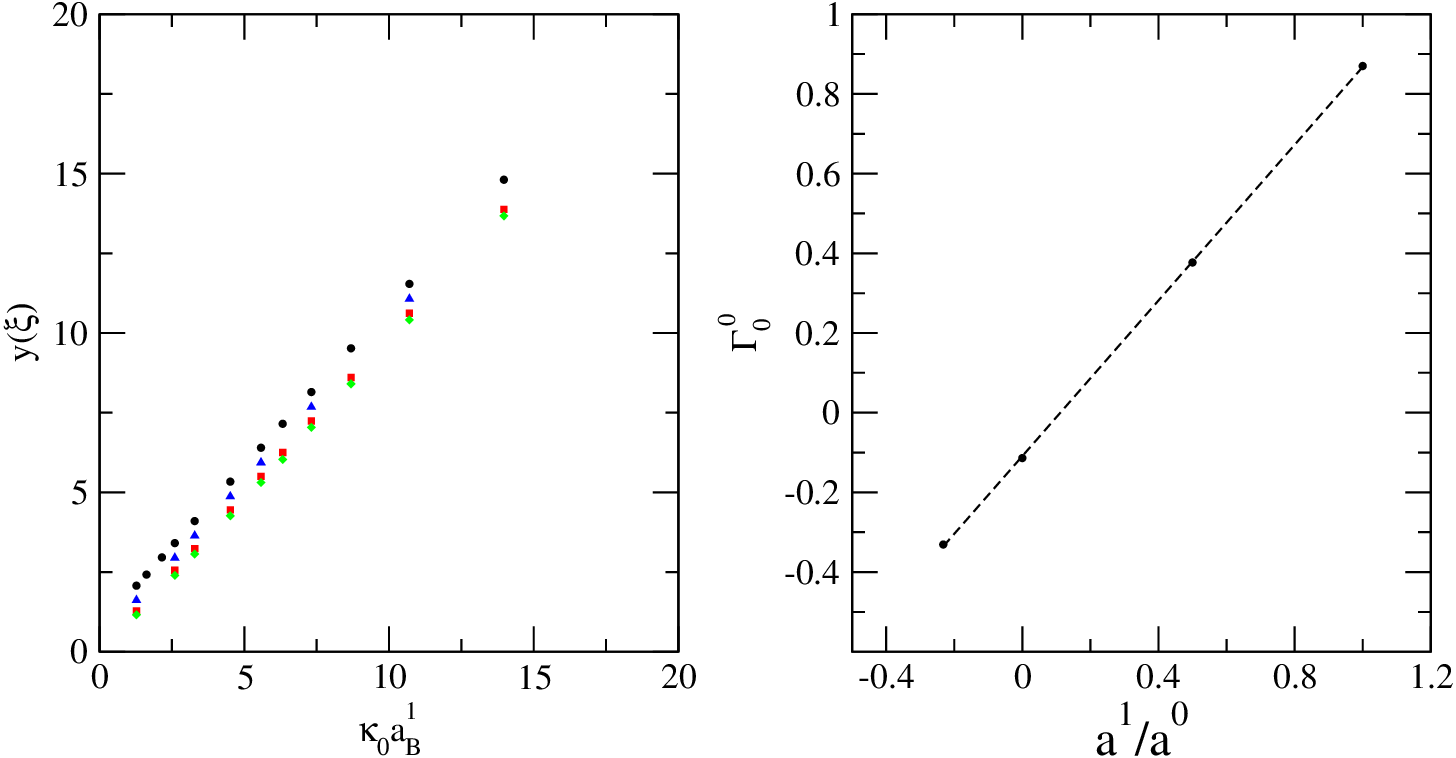}
\end{center}
\caption{(color online). Left panel: the function $y(\xi)$ as a function of the dimensionless
quantity $\kappa_0 a_B^1$ for the four different ratios: $a^0/a_1=1$ (black dots),
$a^0/a_1=2$ (blue triangles), $a^0/a_1=\infty$ (red squares) and 
$a^0/a_1=-4.31$ (green circles). Right panel: the constant term $\Gamma^0_0$ as a function
of the ratio $a^1/a^0$ (black dots). The dashed line is a linear fit to the points.}
\label{fig:gamma0}
\end{figure}

\begin{figure}[hb]
\vspace{3.2cm}
\begin{center}
\includegraphics[width=\linewidth]{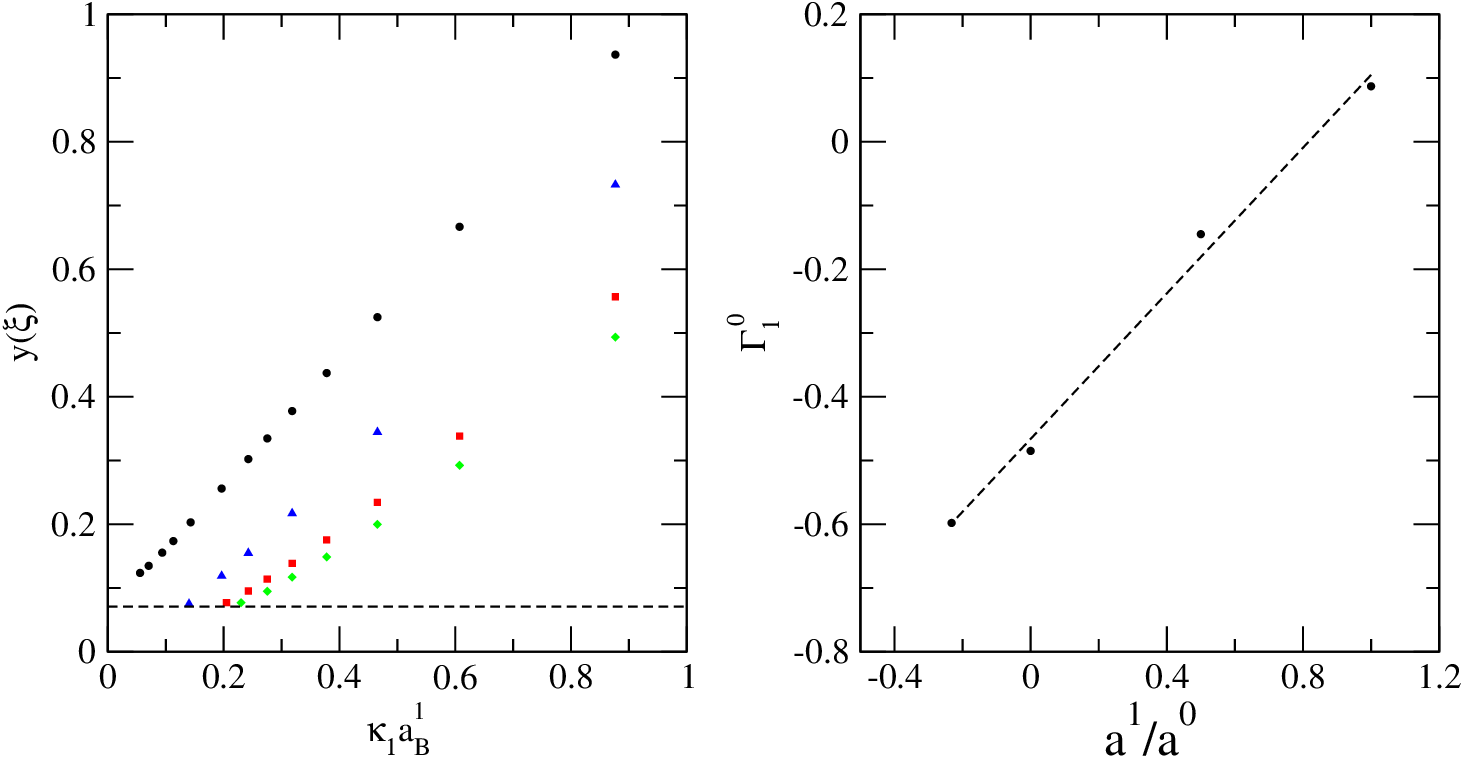}
\end{center}
\caption{(color online). Left panel: the function $y(\xi)$ as a function of the dimensionless
quantity $\kappa_1 a_B^1$ for the four different ratios: $a^0/a_1=1$ (black dots),
$a^0/a_1=2$ (blue triangles), $a^0/a_1=\infty$ (red squares) and 
$a^0/a_1=-4.31$ (green circles). The horizontal dashed line indicates the position of the two-body
threshold. Right panel: the constant term $\Gamma^0_1$ as a function
of the ratio $a^1/a^0$ (black dots).The dashed line is a linear fit to the points.}
\label{fig:gamma1}
\end{figure}

The three-fermion energy spectrum can be analyzed using Eq.(\ref{eq:energyfr}). 
The second row of the equation can be written as
\begin{equation}
  \kappa_n a_B^1+\Gamma_n =y(\xi)
\label{eq:lineary}
\end{equation}
emphasizing the linear relation between the dimensionless variable
$\kappa_na_B^1$ and the universal function $y(\xi)=e^{-\Delta(\xi)/2s_0}/\cos\xi$
for each $n$-level. From the computed values of the energy $E_3^n$ it is possible to evaluate
the angle $\xi$ using Eq.(\ref{eq:energyfrA}) and, accordingly, the specific
value of the function $y(\xi)$. Plotting this function as
a function of the variable $\kappa_na_B^1$ a series of (almost) straight lines with
slope one has to be obtained, from which the shift $\Gamma_n$ can be extracted.
It should be noticed that the shift is not strictly constant but has a dominant
constant term in the expansion in terms of $\kappa_na_B^1$. Explicitly the shift
can be expanded as
\begin{equation}
  \Gamma_n =\Gamma_n^0 + \frac{\Gamma_n^1}{\kappa_n a_B^1}+ \ldots
\label{eq:gammab}
\end{equation}
Close to the unitary limit the shift is dominated by the constant zero-order term $\Gamma_n^0$.
In order to evaluate the quality of the above parametrization in the analysis of the
results, in Fig~\ref{fig:gamma0} (left panel) the function $y(\xi)$ is shown for the ground
state energy in the four cases under consideration, in terms of $\kappa_0a_B^1$.
The linear behavior proposed in Eq.(\ref{eq:lineary}) is well verified. 
In the right panel the values of the zero-order term of the expansion, $\Gamma_0^0$, is
given as a function of the ratio $a^1/a^0$. It can be observed that $\Gamma_0^0$
has an almost linear behavior in terms of the ratio $a^1/a^0$. In Fig.~\ref{fig:gamma1} the
same analysis is done for the excited state $E_3^1$ with similar conclusions. In addition,
in the left panel of the figure the two-body threshold is shown as a dashed line. It
is possible to see the three cases already mentioned in which the excited state disappear 
into the two-body continuum. In conclusion the three-fermion spectrum can be described
by Eq.(\ref{eq:energyfr}) with a shift depending almost linearly with the ratio
$a^1/a^0$.

\section{Hypercentral three-body force}\label{sec:threeb}

\begin{figure}[h]
\vspace{1.2cm}
\begin{center}
\includegraphics[width=\linewidth]{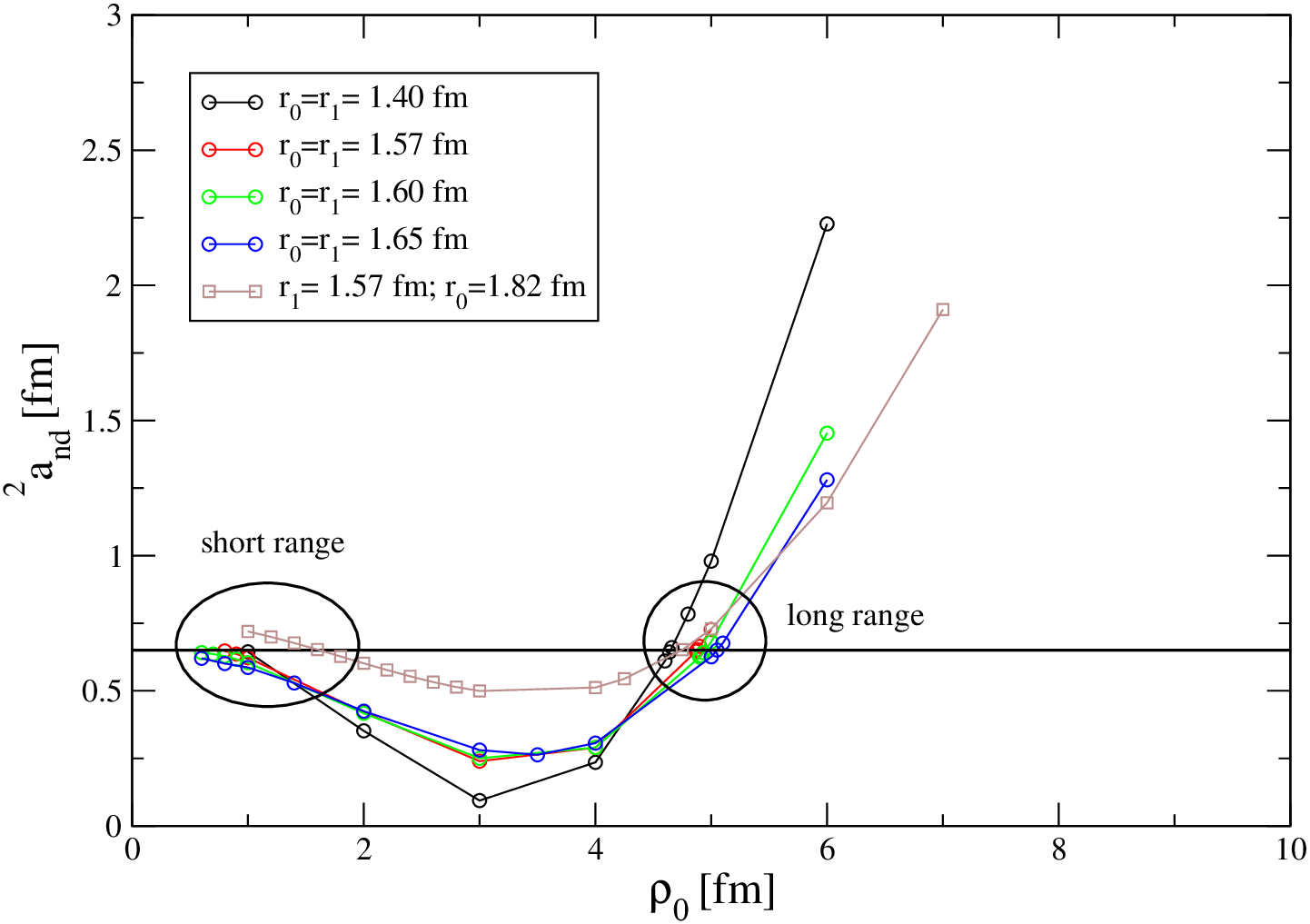}
\end{center}
\caption{(color online). The doublet scattering length $^2a_{nd}$ as a function of
the range $\rho_0$ of the three-body force for different combinations of the
two-body potentials. The horizontal thick (black) line indicated the experimental value
 of $^2a_{nd}$.}
\label{fig:and}
\end{figure}

In this section we discuss the three- and four-fermion systems including two-body
and three-body forces. As in Refs.~\cite{gatto2014,kievsky2013} we propose the
following two-parameter hypercentral three-body force
\begin{equation}
W(\rho)=W_0 e^{-\rho^2/\rho_0^2} \; ,
\label{eq:3bf}
\end{equation}
where $\rho$ is the three-body hyperradius $\rho=2/3(r_{12}^2+r_{23}^2+r_{31}^2)$ and
$r_{ij}$ are the interparticle distances. The three-body force proposed is one of the simplest
ones and, with a proper selection of the parameters $W_0$ and $\rho_0$, we expect an improvement
in the description of the low-energy behavior of the three- and four-fermion systems. 
We make this analysis at the physical point in which $a^0$ and $a^1$ are close to the experimental
values and we fix the two parameters of the three-body force
in order to describe the triton binding energy and the
doublet $n-d$ scattering length $^2a_{nd}$. To this aim, we solve the Schr\"odinger equation 
for three-fermions with total angular momentum $L=0$ and $J^\pi=1/2^+$
using the Gaussian potentials with the parameters given in the first four
rows of Table~\ref{tab:tab1} corresponding to $r_0=r_1=1.40,1.57,1.60,1.65\;$ fm. 
In addition we consider one case in which the singlet state 
parameters are given in the last row of the table ($V_0=-30.932$ MeV and $r_0=1.82\;$fm) and, 
for the triplet state, in the second row ($V_0=-65.74$ MeV and $r_0=1.57\;$fm). This last case,
with different ranges in the two spin states, is analyzed in order to make the optimum choice in
each spin channel. The results, including the hypercentral three-body force, 
are given in Fig.~\ref{fig:and} for $^2a_{nd}$ as a function
of $\rho_0$. At each point of the graph, the strength of the three-body force, $W_0$, 
has been fixed to reproduce the triton
binding energy of $-8.48$ MeV. From the figure we can see two different regions, a short-range
region and a long-range region, in which the experimental value $^2a_{nd}\approx0.65\;$fm
(given by the horizontal thick black solid line)
is reproduced for the different set of Gaussian potentials.

In order to analyze further these
two regions, we calculated the alpha particle ground energy
$E_4^0$, for the different
combinations of Gaussian potentials and three-body forces. The results are
collected in Table~\ref{tab:tab2}. In the upper part of the table the four-body
energies corresponding to the short-range three-body force are given whereas
in the lower part the energies corresponding to long-range three-body force are shown.
The experimental $^4$He ground state energy of $-28.3$ MeV is better describe by the short-range set
of parameters (the case with $r_0=r_1=1.4\;$fm is an exception). In particular the optimum choice,
given by the parameters in the fourth row of the table, produces a result very close to the 
experimental value.

The spin-dependent two-body Gaussian potential plus the hypercentral force can be
consider a potential model that can be used to describe the low-energy $L=0$ states in the two-,
three- and four-fermion system in which the spatial-symmetric state is dominant. In the
particular case of the optimum choice, this potential model describes simultaneously
the $^3$H and $^4$He binding energies and the doublet $n-d$ scattering length $^2a_{nd}$.
Many realistic NN potentials including three-nucleon interactions cannot
describe simultaneously these three quantities~\cite{kievsky2008}. Only recently, in the context of the
chiral effective field theory, three-body forces at next-to-next leading order (N2LO)
can be used to describe the $^3$H binding energy and $^2a_{nd}$ tuning 
the $c_D$ and $c_E$ constants of the (regularized) three-body contact terms~\cite{kievsky2010}.
The accuracy of the description for the three quantities using realistic potential
models is similar to the
accuracy achieved here. A key element in the present analysis was the study of the
observables in terms of the range of the three-body force, similar analyses using realistic
potential models are at present underway~\cite{viviani2013}.

\begin{table}[h]
\begin{tabular}{llll|llc}
 $V_0$[MeV] & $r_0$[fm]    & $V_1$[MeV] & $r_1$[fm] & 
 $W_0$[MeV] & $\rho_0$[fm] & $E_4^0$ \\
\hline
 $-53.255$ & $1.40$ & $-79.600$ & $1.40$ & $215.397$ & $1.00$ & $-23.7$ \\
 $-42.028$ & $1.57$ & $-65.750$ & $1.57$ & $532.046$ & $0.80$ & $-26.9$ \\
 $-40.413$ & $1.60$ & $-63.712$ & $1.60$ & $2979.09$ & $0.60$ & $-27.3$ \\
 $-30.932$ & $1.82$ & $-65.750$ & $1.57$ & $17.265 $ & $1.60$ & $-28.4$ \\
\hline
 $-53.255$ & $1.40$ & $-79.600$ & $1.40$ & $6.1172$  & $4.64$ & $-27.5$  \\
 $-42.028$ & $1.57$ & $-65.750$ & $1.57$ & $3.7708$  & $4.87$ & $-30.4$  \\
 $-40.413$ & $1.60$ & $-63.712$ & $1.60$ & $3.3898$  & $4.93$ & $-30.4$  \\
 $-37.900$ & $1.65$ & $-60.575$ & $1.65$ & $2.7947$  & $5.05$ & $-30.3$  \\
 $-30.932$ & $1.82$ & $-65.750$ & $1.57$ & $2.1216$  & $4.75$ & $-30.7$  \\
\hline
\end{tabular}
\caption{The four-body ground state energy for different set of two-body and three-body
forces.}
\label{tab:tab2}
\end{table}

\section{Conclusions}

In the present work we have studied two-, three- and four-fermion systems with $1/2$ spin-isospin
symmetry. The study focussed on the dynamics of the $L=0$ states close to the 
unitary limit in which a universal behavior is expected. For $1/2$ spin-isospin fermions,
the $L=0$ state has a large spatially symmetric component making this system similar to the
bosonic system. In the latter the two-body system is characterized by the scattering length
$a$ whereas, in the fermion case the system, is characterized by two scattering lengths, 
$a_0$ and $a_1$, in the two spin states $S=0,1$. For large values of these quantities the
spectrum of the system has been calculated using a potential model. We have selected a
Gaussian form for the two-body potential. Having in mind the nuclear system, in a first
step, the range and strength of the potential in each spin channel have been fixed to reproduce 
$a_0$ and $a_1$, the deuteron energy and effective range parameters. In a second step
the strength of the potentials has been varied to explore the dynamics close to the unitary
limit. To organize the study the spectrum, the variation of the Gaussian strengths have not
been done independently but along lines where the ratio $a_0/a_1$ has been kept fixed.

In the two-body system potentials with variable strength define a particular path to reach the unitary
limit. Along this path we have shown that the length $r_B=a-a_B$ remains almost constant.
Furthermore we have derived a universal relation for the effective range that connects
linearly the regions with positive and negatives values of $a$. This relation, well verified
by the Gaussian potential is also verified by a much realistic potential consisting 
in a sum of two Yukawians. We can conclude that a two-parameter potential has the
essential ingredients to describe the dynamics close to the unitary limit. We extend the
analysis to calculate the three-fermion spectrum for selected values of the ratio $a_0/a_1$.
In this study we used Gaussians of equal range, accordingly the case $a_0/a_1=1$ represents
the bosonic case. We have analyzed other three cases and, in particular
$a_0/a_1=-4.31$ which corresponds to the nuclear physics case. Two important conclusions
can be extracted. In first place the energy spectrum can be described with the modified
radial low of Eq.(\ref{eq:energyfr}) in which the shift, $\Gamma_n$, depends on the ratio 
$a_0/a_1$. For the ground state, the dominant constant term $\Gamma_0^0$ varies almost
linearly from 0.87 at $a_0/a_1=1$ to -0.23 at $a_0/a_1=-4.31$. A second result regards
the first excited state. The zero-range theory for bosons, given by Eq.(\ref{eq:energyzr}), 
predicts that the energy states are absorbed into the two-body continuum at specific values of $a$.
The description using potential models shows that this is the case for the higher excited states.
However the ground and first excited state remains always below the two-body energy as the
values of $a$ increases. This situation changes in the case of fermions, as the ratio
$a_0/a_1$ increases making the singlet potential weaker than the triplet one, the first excited
state is absorbed into the two-body continuum. In particular, at the physical point,
$a_0/a_1=-4.31$ and $a_1\approx 5.4\;$fm there is only one bound state in the spectrum
in agreement with the experimental observation.

In the last part of the study we have included an hypercentral two-parameter
three-body force in order to study correlations between the low energy observables in
the three- and four-fermion system. The correlation between the triton binding energy
and the doublet scattering length is the Phillips line~\cite{phillips}. Here we have shown
that in order to describe both quantities simultaneously a detailed analysis of their
dependence on the range of the hypercentral three-body force was needed. 
Two well separated regions, one short-range and one long-range, were observed both giving
a good description of the observables. In order to analyze further the impact of these
two regions in the description of the observables, the four-body ground state energy was 
calculated. The conclusion was that the short-range three-body force gives a better
description of the $^4$He ground state energy allowing for a simultaneous description
of the three quantities. In particular the optimum choice, with different ranges in
the Gaussian two-body potentials, describes the three observables with reasonable
accuracy.

From this study we have observed that the three and four-nucleon systems are
inside the Efimov window in which the structure of the system is governed by
a few control parameters as the two-body energies and scattering lengths. 
Introducing this minimal
information in the construction of the potentials, the correlations in the 
low-energy observables can be properly described. This happens for the $L=0$
state which is mostly spatially symmetric. For $A>4$ the symmetry of the
spatial part changes and the present analysis has to be modified accordingly.
Studies along this line are at present underway.

%\bibliographystyle{fewbodysystems}
%\bibliography{FewBodyBiblio}

\end{document}